УДК 004.891

# Некоторые результаты экспериментальной проверки модели количественной оценки инновационности объекта

# Some Results of Experimental Check of The Model of the Object Innovativeness Quantitative Evaluation

**Иванов В.К.,**
кандидат технических наук, доцент, начальник управления информационных ресурсов и технологий, Тверской государственный технический университет, Тверь, Россия, mtivk@tstu.tver.ru

**Ivanov V.K.,**
Ph. D. (Engineering) sciences, Associate professor, Head of Department of Information Resources and Technologies, Tver State Technical University, Tver, Russia, mtivk@tstu.tver.ru

**Аннотация**. В статье представлены результаты экспериментов, которые были проведены для подтверждения основных идей предлагаемого подхода к определению инновационности объектов. Этот подход основан на предположении об адекватности отображения жизненного цикла продуктов, описания которых размещены в различных хранилищах данных. Предложенная формальная модель позволяет вычислить количественное значение аддитивного оценочного критерия инновационности объектов. Полученные данные экспериментов дают возможность оценить корректность принятого подхода.

**Ключевые слова**: хранилище данных, инновационность, аддитивный критерий, функция полезности, поисковый запрос

**Abstract**. The paper presents the results of the experiments that were conducted to confirm the main ideas of the proposed approach to determining the objects innovativeness. This approach assumed that the product life cycle of whose descriptions are placed in different data warehouses is adequate. The proposed formal model allows us to calculate the quantitative value of the additive evaluation criterion of objects innovativeness. The obtained experimental data make it possible to evaluate the adopted approach correctness.

**Keywords**: Data Warehouse, Innovation, Additive Criterion, Utility Function, Search Query.

## Введение

Условия современного развития производственных процессов предполагают очевидный интерес к объектам (продуктам и технологиям), обладающим значимым инновационным потенциалом. Отсюда возникает необходимость в количественной оценке такого потенциала. Мы предлагаем подход, который позволяет оценить инновационный потенциал объектов с помощью индекса инновационности, модель вычисления которого основана на таких показателях инновационности, как новизна, востребованность и имплементируемость объекта.

Исходными данными для вычисления показателей инновационности объекта служат его описания, в том или ином виде хранящиеся в различных гетерогенных базах данных. Отметим, что указанные показатели вычисляются не для самого объекта, а для его лингвистической модели. Лингвистическая модель объекта используется для генерации множества запросов к доступным базам данных. Характеристики множеств релевантных описаний объекта, найденных при выполнении запросов, используются при вычислении показателей инновационности. Вычисленные показатели сводятся к глобальному аддитивному критерию — индексу инновационности объекта.

При вычислении показателей инновационности используются формальные выражения, основанные на ряде предположений, которые описаны далее в статье. Например, предположение о том, что имплементируемость или реализуемость продукта зависит от величины периода восстановления спроса на его улучшенную модификацию. Или принятие в качестве одного из параметров востребованности продукта количества запросов пользователей к базам данных с упоминанием характеристик этого продукта. Естественно, эти предположения требуют экспериментального подтверждения с помощью соответствующих методик и инструментов для измерения, обработки и сравнения значений обсуждаемых показателей.

Цель статьи — экспериментально подтвердить некоторые предположения, использованные при формализации модели для вычисления аддитивного оценочного критерия инновационности объектов. Некоторые результаты экспериментальной проверки этой модели, впервые представленные в настоящей статье, дают возможность в определенной степени оценить корректность принятого подхода. Настоящая статья является продолжением работ автора и его коллег по обсуждаемой тематике [1, 2 и др.], которые направлены на развитие методов и средств повышения эффективности информационного поиска в различных прикладных областях (см., например, [3]).

**Лингвистическая модель объекта**

Лингвистическая модель объекта может быть представлена как:

$$LM = \{A_s, A_c, A_r, M, C\} \quad (1)$$

где $A_s$, $A_c$ и $A_r$ — множества или классы архетипов объекта, определяющих соответственно структуру, условия применения и результаты функционирования этого объекта, при этом $|A_s| = |A_c| = |A_r|$. Архетип объекта $a \in A$ — это концепции предметной области для рассматриваемого объекта, которые реализуются термами, определяющими ключевые свойства объекта, и группируются в классы $A_s$, $A_c$ и $A_r$;

$M$ — множество термов-маркеров $m \in M$, задающее область определения архетипов объекта.

$C$ — дополнительные локальные ограничения: умолчания, синонимы термов, веса термов, предельное количество запросов, количество термов в запросе и т.п. Эти ограничения используются в алгоритмах генерации поисковых запросов.

Примеры архетипов и маркера приведены в таблице 1.

*Таблица 1*

**Примеры архетипов и маркера для объекта «смартфон»**

| Архетипы $A_s$ | Архетипы $A_c$ | Архетипы $A_r$ | Маркер $M$ |
|---|---|---|---|
| *камера* | *NFC* | *производительность* | *смартфон* |
| *экран* | *4G* | *аккумулятор* | *Galaxy S* |

Примечание к таблице 1: маркеры $m \in M$ ограничивают множество анализируемых смартфонов только устройствами одного семейства *Samsung Galaxy S*.

Лингвистическая модель (1) используется как поисковый паттерн для генерации набора запросов для поиска информации о потенциально инновационном объекте. Поисковые запросы — логические выражения, где множества операндов есть различные комбинации термов $a \in A$ и $m \in M$. Релевантные документы, найденные после исполнения всех сгенерированных запросов, используются для вычисления отдельных показателей инновационности объекта и индекса инновационности объекта в целом.

## Модель вычисления индекса инновационности объекта

Количественная оценка показателей инновационности объекта основывается на предположении об адекватности отображения жизненного цикла продуктов в виде их описаний, размещенных в различных хранилищах данных [1].

Показатель технологической новизны $Nov$ вычисляется следующим образом:

$$Nov = 1 - 1/N \sum_{k=1}^{N} f_n^{01}(R_k,...) \qquad (2)$$

где $N$ – общее количество выполненных запросов; $R_k$ — число документов, найденных в результате выполнения $k$-го запроса; $f_n^{01}(R_k,...)$ — вариативная функция, нормирующая значение $R_k$ на диапазон [0;1].

Показатель востребованности $Dem$ вычисляется следующим образом:

$$Dem = 1/S \sum_{k=1}^{S} f_n^{01}(F_k,...) \qquad (3)$$

где $S$ — общее количество выполненных запросов; $F_k$ – частота выполнения $k$-го запроса; $f_k^{01}(F_k,...)$ – вариативная функция, нормирующая значение $F_k$ на диапазон [0;1].

Показатель имплементируемости $Imp$ вычисляется следующим образом:

$$Imp = 1 - 1/2 f_n^{01}(\Delta t_N(Nov(t)) + \Delta t_D(Dem(t))) \qquad (4)$$

где $Nov(t)$ — функция, показывающая зависимость $Nov$ от времени на временном интервале $[t_0; t_m]$; $Dem(t)$ — функция, показывающая зависимость $Dem$ от времени на том же $[t_0; t_m]$; $\Delta t_N$ и $\Delta t_D$ — средние расстояния между двумя последовательными точками временного ряда $t_i, t_{i+1} \in [t_0; t_m]$ локальных максимумов функций $Nov(t)$ и $Dem(t)$; $f_k^{01}$ – вариативная функция, нормирующая значение $\Delta t$ на диапазон [0;1]. При этом $Nov$, $Dem$ и $Imp$ рассчитываются для точки $t_{m+1}$.

Индекс инновационности $Ix$ имеет вид аддитивного критерия:

$$Ix = w_{Nov}Nov + w_{Dem}Dem + w_{Imp}Imp \qquad (5)$$

где $w_{Nov}$, $w_{Dem}$, $w_{Imp}$ — весовые коэффициенты для $Nov$, $Dem$, и $Imp$ соответственно и $w_{Nov} + w_{Dem} + w_{Imp} = 1$.

В случае неполной и неточной информации об объектах вводятся нечеткие показатели вероятности того, что объект обладает новизной, востребованностью и имплементируемостью. Для вычисления указанных вероятностей при-меняется теория свидетельств Демпстера-Шафера [4]. Определяются базовые вероятности $m$ попадания результатов измерения $Nov$, $Dem$, и $Imp$ в $k$-й интервал значений $A_k$; результаты из различных источников рекурсивно комбинируются по парам источников: из двух источников свидетельств образуется один условный источник, свидетельства которого комбинируются с очередным фактическим источником.

Рассчитываются функция доверия $Bel(A) = \sum_{A_k : A_k \subseteq A} m(A_k)$ и функция правдоподобия $Pl(A) = \sum_{A_k : A_k \cap A} m(A_k)$, которые определяют верхнюю и нижнюю границу вероятности обладания объектом свойства, заданного соответствующим показателем. Тогда выражение (5) приобретает вид мультипликативного оценочного критерия:

$$Ix = [Bel_{Nov}(A), Pl_{Nov}(A)]^{w_{Nov}} * [Bel_{Dem}(A), Pl_{Dem}(A)]^{w_{Dem}} \\ * [Bel_{Imp}(A), Pl_{Imp}(A)]^{w_{Imp}} \quad (6)$$

которое сводится логарифмированием $Ix$ к аддитивному критерию:

$$\ln Ix = w_{Nov} * \ln([Bel_{Nov}(A), Pl_{Nov}(A)]) + w_{Dem} * \\ \ln([Bel_{Dem}(A), Pl_{Dem}(A)]) + w_{Imp} * \ln([Bel_{Imp}(A), Pl_{Imp}(A)]) \quad (7)$$

Так как $\ln()$ возрастающая функция, рассуждения, касающиеся $Ix$, справедливы для $\ln(Ix)$.

**Методика проведения экспериментов**

В ходе проведения экспериментов планировалось получить данные, подтверждающие следующие предположения, использованные при формулировке выражений (2), (3) и (4):

— необходимость использования многих источников информации об объектах для сбора исходных данных при вычислении показателей инновационности.

— применимость предложенной модели вычисления индекса инновационности объекта для различных типов объектов.

— зависимость показателя новизны объекта от времени есть очевидное уменьшение его значения. Следовательно, количество информации об объекте, найденной в базах данных, должно со временем увеличиваться.

— цикличность изменения показателей инновационности вследствие улучшений конструкции, технологии использования, эксплуатационных характеристик объектов.

— использование в качестве параметров востребованности объекта различных характеристик пользовательского доступа к информации о нем.

— зависимость показателя имплементируемости объекта от величины периода создания его улучшенной модификации и соответствующего восстановления спроса.

В качестве объектов с инновационным потенциалом были отобраны 37 продуктов и технологий. В их число входили: смартфоны популярных моделей известных производителей, способы получения новых материалов, космические аппараты, медицинские технологии, технологии приготовления продуктов питания, технические устройства.

После этого экспертами были сформированы лингвистические модели для каждого объекта. Эти модели были использованы для генерации поисковых запросов.

В качестве источников данных об объектах использовались следующие хранилища данных: ACM Digital Library (https://dlnext.acm.org), AliExpress (https://aliexpress.com), Google (https://google.com), Google Patents (https://patents.google.com), Google Scholar (https://scholar.google.ru), IEEE Explore Digital Library (https://ieeexplore.ieee.org), ЕГИСУ НИОКТР (https://rosrid.ru), Научная электронная библиотека (https://elibrary.ru), поисковая система ФИПС (https://www1.fips.ru/iiss), Яндекс (https://yandex.ru), а также базы данных системы электронного обучения и учебно-методических комплексов ТвГТУ (http://elearning.tstu.tver.ru).

Показатели $Nov$ и $Dem$, измеренные по результатом выполнения совокупности сгенерированных поисковых запросов, усред-нялись среднеарифметическим и медианным значениями, а также нормализовались на диапазон [0;1] функциями $f_n^{01}$ для линейного $f_{nk}^{01} = \frac{F - \min F}{\max F - \min F}$ и экспоненциального $f_{ne}^{01} = 1 - \exp(1 - \frac{F}{\min F})$ нормирования. Здесь $F$ использовано для обозначения $Nov$ или $Dem$ в соответствующих случаях.

### Результаты экспериментов

На рис. 1 представлены средние значения показателя $Nov$ для всех исследованных объектов, вычисленные на основании информации, накопленной за 23 года во всех использованных источниках данных. По совокупности результатов выполнения запросов вычислялись:

среднеарифметические значения $Nov\_1\tilde{c}$ и $Nov\_4\tilde{c}$, нормированные функциями $f_{nk}^{01}$ и $f_{ne}^{01}$ соответственно;

медианные значения $Nov\_1m$ и $Nov\_4m$, нормированные функциями $f_{nk}^{01}$ и $f_{ne}^{01}$ соответственно.

Для каждого ряда значений показаны линии трендов (линейная аппроксимация).

На рис. 2 показан значения показателя $Nov$ для двух исследованных объектов из области современной медицины: ген-активированного материала для регенерации тканей и технологии лечения пародонта. Вычисления выполнены на основании информации, накопленной за 21 год в двух источниках данных: elibrary.ru и Google Scholar. По совокупности результатов выполнения запросов вычислялись среднеарифметические значения $Nov\_4\tilde{c}$, нормированные функцией $f_{ne}^{01}$. Для каждого ряда значений показаны линии трендов (полиномиальная аппроксимация).

Рис. 3 иллюстрирует динамику изменения средних значений показателя $Dem$ для ген-активированного материала для регенерации тканей и технологии лечения пародонта. Вычисления выполнены на основании информации, накопленной за 20 лет в двух источниках данных: elibrary.ru и Google Scholar. По совокупности результатов выполнения запросов вычислялись среднеарифметические значения $Dem\_4\tilde{c}$, нормированные функцией $f_{ne}^{01}$. Для каждого ряда значений показаны линии трендов (полиномиальная аппроксимация).

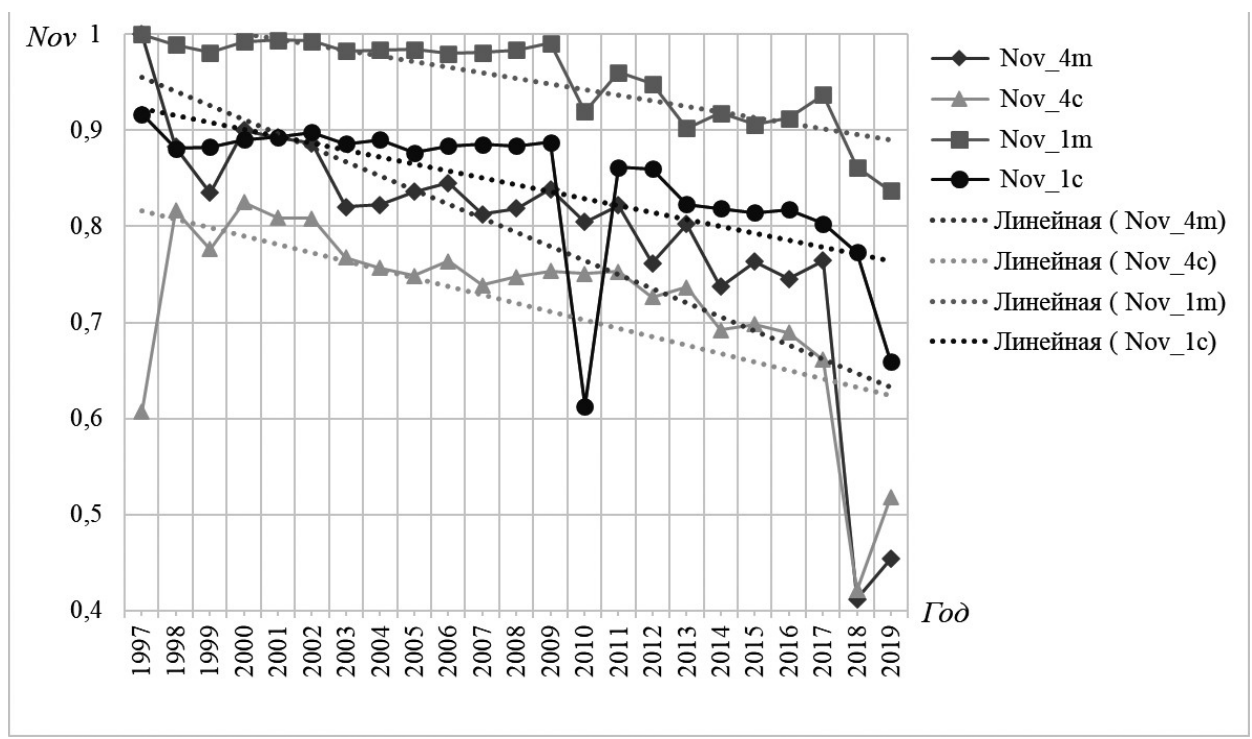

*Рис. 1 Средние значения показателя $Nov$ по годам (все исследованные объекты, все использованные источники данных)*

На рис. 4 представлены средние значения показателя $Dem$ для двух исследованных объектов, вычисленные на основании информации, накопленной за 20 лет в двух источниках данных. По совокупности результатов выполнения запросов вычислялись:

среднеарифметические значения $Dem\_1\tilde{n}$ и $Dem\_4\tilde{n}$, нормированные функциями $f_{nk}^{\overline{01}}$ и $f_{ne}^{01}$ соответственно;

медианные значения $Dem\_1m$ и $Dem\_4m$, нормированные функциями $f_{nk}^{01}$ и $f_{ne}^{01}$ соответственно.

Для каждого ряда значений показаны линии трендов (линейная аппроксимация).

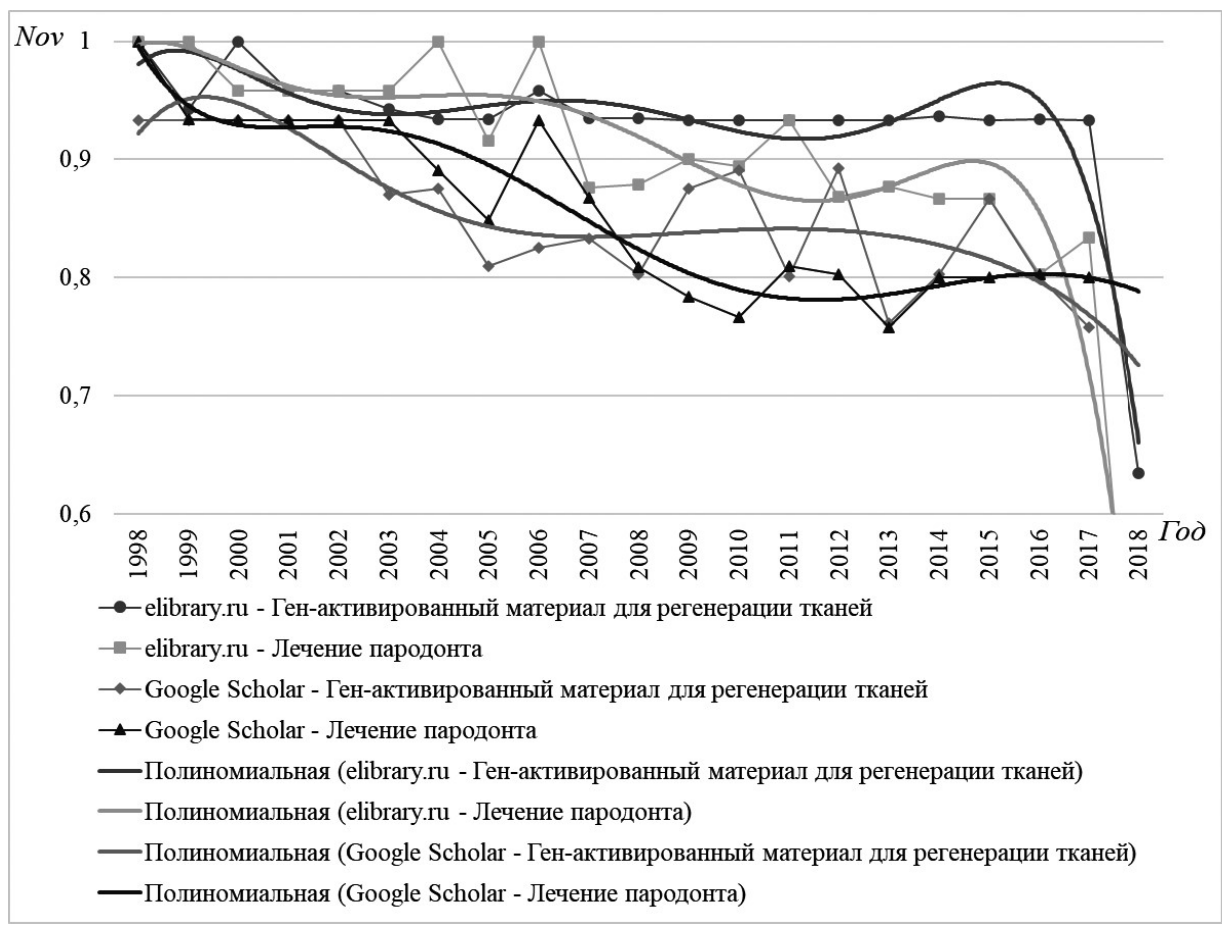

*Рис. 2 Значения показателя $Nov$ по годам для двух объектов и двух источников данных*

На рис. 5 показано сравнение вычисленных показателей $Nov$ и $Dem$ для объектов, отобранных экспертами как инновационные, и случайно отобранных объектов. Для проведения экспериментов использовались описания изобретений из базы данных ФИПС (http://www1.fips.ru/iiss): 10 самых значимых изобретений 2017 года, выбранных экспертами «Роспатента» [5]. Вторая группа — 10 случайно выбранных в базе данных Роспатента. По совокупности результатов выполнения запросов вычислялись:

средние значения $Nov\_4\tilde{n}$ и $Dem\_4\tilde{n}$, нормированные функцией 32

медианные значения $Nov\_4m$ и $Dem\_4m$, нормированные функцией $f_{ne}^{01}$.

### Обсуждение результатов

Очевидно, что использование многих источников информации об объектах для сбора исходных данных при вычислении показателей инновационности позволяет получить более адекватные значения. Например, на рис. 2

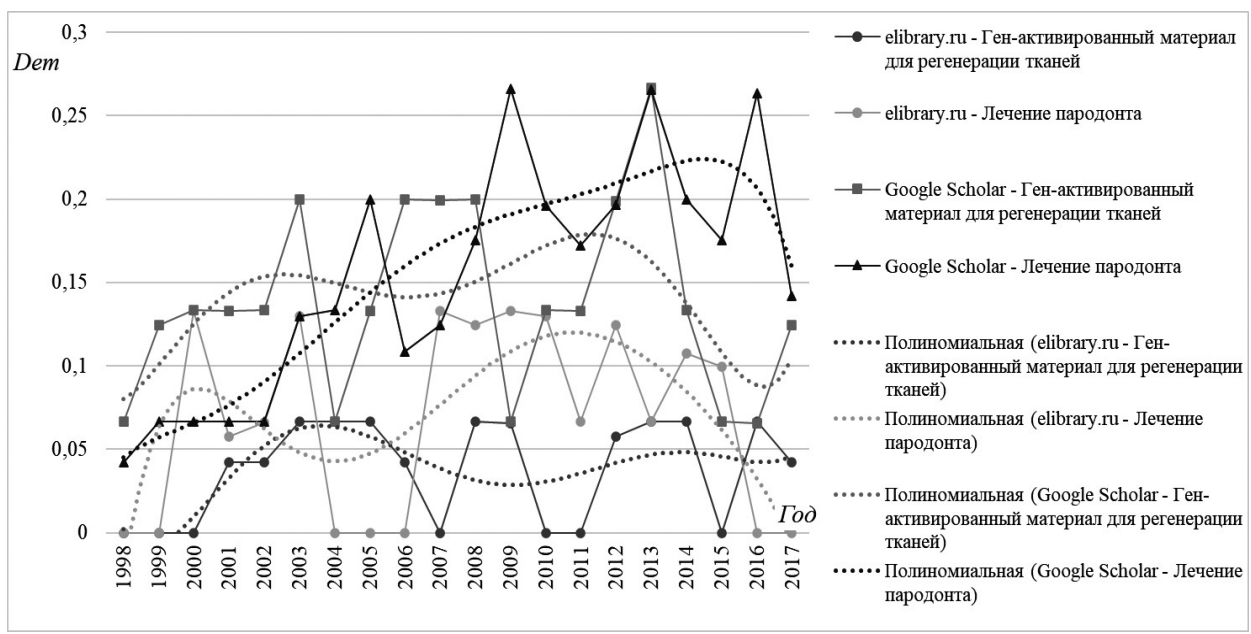

*Рис. 3 Средние значения показателя Dem по годам
для двух объектов и двух источников данных*

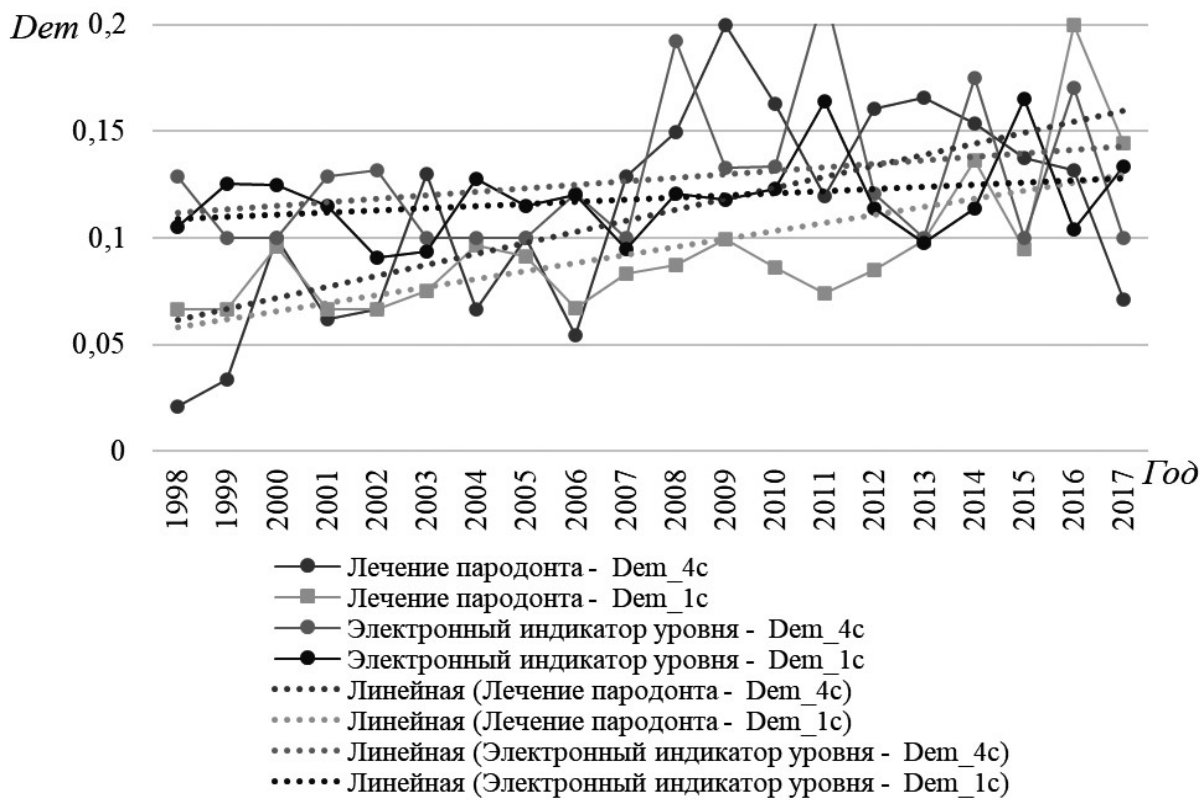

*Рис. 4 Значения показателя Dem по годам
(два исследованных объекта, два источника данных)*

для технологии лечения пародонта данные из баз данных elibrary.ru и Google Scholar дают похожие результаты по виду аппроксимирующей кривой, но абсолютные значения показателя $Nov$ отличаются. Аналогичная ситуация с другим объектом, данные о котором представлены на этом рисунке. Рис. 3 дает сходную картину для показателя $Dem$: похожие аппроксимирующие кривые, но отличающиеся (иногда существенно) значения показателей. Простое усреднение измеренных значений может улучшить окончательный расчет. Предлагаемое использование таких методов, как теория свидетельств Демпстера-Шафера [4], позволяет еще более обосновано выполнить комбинирование отличающихся результатов.

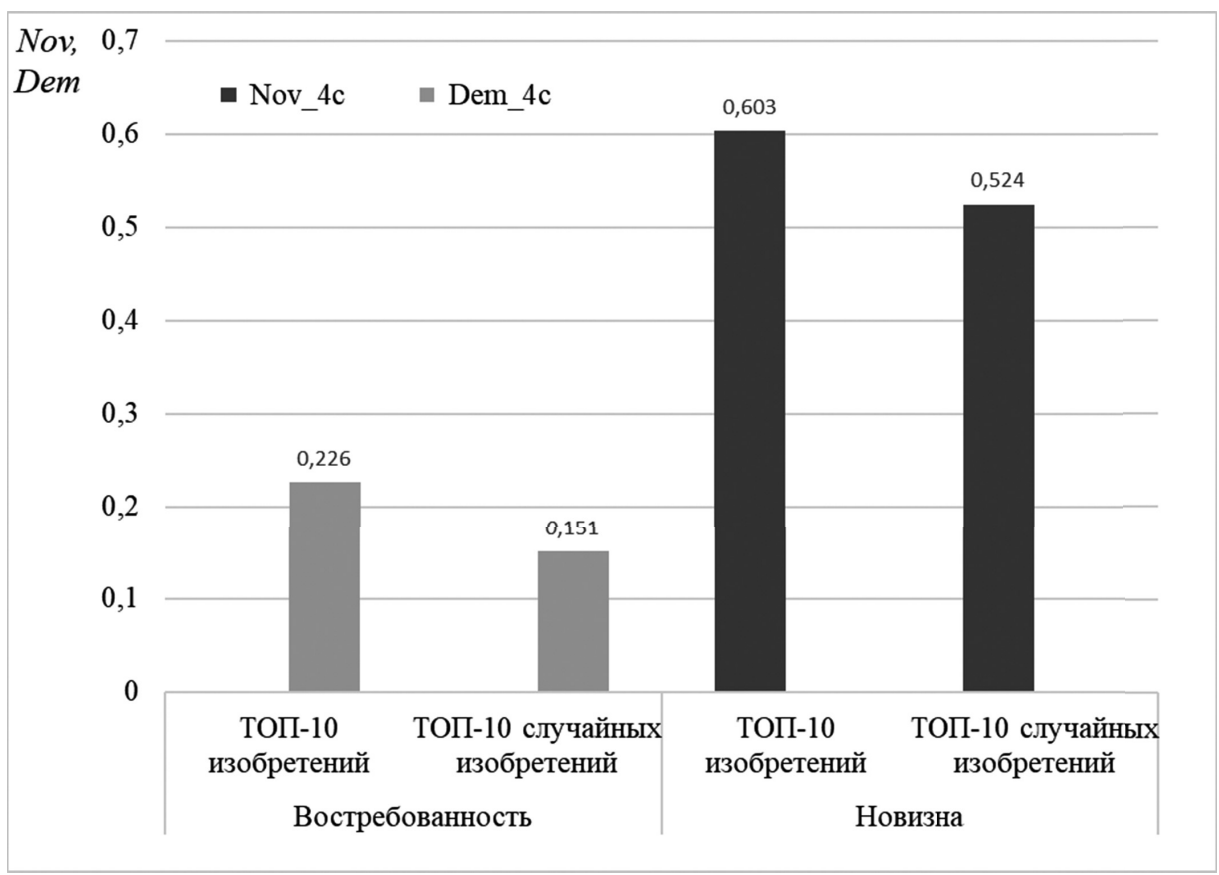

*Рис. 5 Сравнение вычисленных показателей $Nov$ и $Dem$ для объектов, отобранных экспертами как инновационные, и случайно отобранных объектов*

Как следует из графика на рис. 1 зависимость показателя новизны объекта от времени есть очевидное уменьшение его значения — линейная аппроксимация рядов экспериментальных значений есть монотонно убывающая функция. Аналогичная ситуация наблюдается практически для всех исследованных объектов (например, рис. 2). При этом характер тренда не зависит ни от вида используемой нормирующей функции, ни от используемого источника данных. Достаточно большой временной интервал, на котором измерялись показатели времени, позволяет говорить о приемлемой точности результатов.

На рис. 2, 3 и 4 показана цикличность изменения показателей $Nov$ и $Dem$ для различных объектов в течение более, чем 20-летнего периода. Отметим на рис. 2 приблизительные локальные экстремумы $Nov$ для технологии лечения пародонта: 2005 и 2015 гг. (elibrary.ru) и 2003 и 2016 гг. (Google Scholar). На рис. 4 отметим приблизительные локальные экстремумы $Dem$ для того же объекта: 2000 и 2012. Вероятная интерпретация этих данных такова: падение спроса после достижения его макси-мума вызывает улучшения конструкции, технологии использования, эксплуатационных характеристик объекта. Следствием этого является улучшение показателя новизны объекта, что мы и видим на графиках.

Цикличность изменения показателей $Nov$ и $Dem$ используется при вычислении показателя имплементируемости объекта $Imp$ в соответствии с (4). Рассмотрим расстояния между двумя последовательными точками временного ряда локальных максимумов функций $Dem(t)$ для объекта «ген-активированного материала для регенерации тканей». Как следует из рис. 3, эти точки соответствуют 2011 и 2003 гг. Аналогичные расстояния для объекта «электронный индикатор уровня» как следует из рис. 4 соответствуют точкам 2011 и 2002 гг. Как видно, период восстановления спроса вследствие различных улучшений обоих объектов в данном примере достаточно большой (8-9 лет). Нормализованные значения $Imp$ = 0,42 в первом случае и $Imp$ = 0,47 — во втором.

Проведенные эксперименты показали совпадение результатов оценки инновационности объектов с помощью вычисленных показателей с экспертными оценками тех же объектов. Рис. 5 хорошо иллюстрирует это. Среднее значения вычисленных показателей $Dem$ для всех объектов, отобранных экспертами как инновационные, превышает аналогичное значение для случайно отобранных объектов на 49%. Аналогичное превышение среднего значения вычисленного показателя $Nov$ — 15%.

Рассмотрим применимость предложенной модели вычисления индекса инновационности объекта для различных типов объектов. Данные экспериментов, некоторые из которых представлены в настоящей статье, показывают сходную динамику поведения измеряемых показателей для различных объектов и источников информации о них. Это касается снижения значений новизны $Nov$ (рис. 1 и 2) и повышения значений востребованности $Dem$ (рис. 4) со временем, цикличности изменений значений $Nov$ $Dem$ (рис. 2 и 3).

Отметим использование в выполненных экспериментах различных показателей в качестве параметров востребованности объекта. Так, базы данных ACM Digital Library, Google Patents, Google Scholar, IEEE Explore Digital Library и НЭБ использовались для получения количества цитирований материалов об объекте, Яндекс и AliExpress — для получения количество запросов с характеристиками объекта и количество продаж объекта. Во всех случаях полученные данные позволили выполнить требуемые вычисления.

В целом анализ представленных результатов показывает, что основные предположения, использованные при формулировке выражений (2), (3) и (4), подтверждены.

## Заключение

Приведенные в статье результаты исследований являются частью проекта РФФИ «Организация и поддержка хранилища данных на основе интеллектуализации поискового агента и эволюционной модели отбора целевой информации» [6]. Они будут использованы в разработке программного обеспечения для селекции информации об инновационных объектах. В целом автоматизация процессов количественной оценки ключевых характеристик современных продуктов и технологий (включая их динамику) должна найти применение в таких областях, как: выбор направлений бизнеса и объектов инвестиций, экспертиза проектов, проведение сравнительного анализа различных продуктов.

## Благодарности

**Конфликт интересов**
Конфликт интересов отсутствует.